\documentstyle[11pt]{article}

\begin{document}
\centerline {\bf ON THERMODYNAMICS OF MULTISPECIES ANYONS}
\vskip 1cm
\centerline {{\bf Serguei B.~Isakov}$^*$,}
\centerline{{\bf Stefan Mashkevich}$^{**,\dagger}$ \footnote{\it
email: mash@phys.unit.no} and
{\bf St\a'ephane Ouvry}$^\dagger$ \footnote{\it  and
LPTPE, Tour 12, Universit\'e Paris 6 / email: OUVRY@IPNCLS.IN2P3.FR}}
\vskip 1cm
\centerline{$^*$ Medical Radiology Research Center, Obninsk,
Kaluga Region 249020, Russia}
\centerline{$^{**}$ Institute for Theoretical Physics, Kiev 252143, Ukraine}
\centerline{{$^\dagger$ Division de Physique Th\'eorique%
\footnote{\it Unit\a'e de Recherche  des
Universit\a'es Paris 11 et Paris 6 associ\a'ee au CNRS}, IPN,
  Orsay Fr-91406}}
\vskip 1cm
{\bf Abstract:}
We address the problem of multispecies anyons, i.e.~particles
of different species whose wave function is subject to
anyonlike conditions. The cluster and virial coefficients are
considered. Special attention is paid to the case of anyons in the
lowest Landau level of a strong magnetic field, when it is possible
(i) to prove microscopically the equation of state,
 in particular in terms of Aharonov-Bohm charge-flux composite systems,
and (ii) to formulate the problem in terms of
single-state statistical distributions.
\vskip 1cm

IPNO/TH 95-21

March 1995

\vfill\eject

\section{Introduction}

Considering quantum mechanical indistinguishable particles, one uses to
speak about their ``statistical'' (or ``exchange'') interaction.
It has no classical analogue, since there is no classical interaction
force, but it influences the many-particle spectrum.
Under usual circumstances, with  bosons and fermions only,
the rules of combining single-particle states for indistinguishable
particles are affected, while particles of different species
remain independent. For anyons
\cite{Leinaas77,Wilczek82}, however,
this interaction is nontrivial, since it leads to
the absence of single-particle states,
and even more,  to the possibility that particles of different species
can become mutually ``entangled''; hence the problem of multispecies
anyons demands special consideration.

It is also believed that this problem may be of direct physical
interest, particularly in connection with the $t-J$ model,
which might be relevant to the high-$T_c$
superconductivity. The question about statistics of excitations
therein, spinons and holons, is still unsettled, and it is anticipated
that their ``mutual statistics'' may be nontrivial \cite{Mor}.
Another possible application is the layered Hall media where FQHE
quasiparticles in different layers can be viewed as anyons with mutual
statistical interactions \cite{Wilczek92}. Finally, the same
situation arises in the usual quantum Hall liquid: In a state
where exactly $m$ Landau levels are filled, the excitations are
characterized by ``charge vectors'' of dimension $m$ and there
are different nontrivial phase factors associated with interchanging
different sorts of excitations \cite{Froelich91}.

In this paper we address certain points concerning thermodynamics
of multispecies anyons, by starting from microscopic considerations on
the wave functions.
For the sake of simplicity, we do the calculations for two species
only, but a generalization to more than two species would
be straightforward. The fugacity expansion of the thermodynamic
potential now involves two fugacities, and the equation of state,
respectively, two densities; hence,
the cluster and virial coefficients are
labeled by two indices. The mixed virial coefficient of the lowest
order, $a_{11}$, is trivially calculable, but exact expressions for
higher-order ones are  certainly not easier to obtain than
in the one-species case. To proceed further,
we investigate the case when all the particles are in the lowest Landau
level of a strong
magnetic field, and, using microscopic arguments as in \cite{Dasnieres94a},
we obtain the equation of state which coincides with the one
conjectured in \cite{Dasnieres94b} and derived
in \cite{Wu94} from a definition of fractional statistics \`a la
Haldane \cite{Haldane91} (exclusion statistics),
using a state-counting argument.
A mean-field interpretation of this equation is also proposed.
Also, in this (and only in this) case it becomes possible
to interpret the system
in terms of statistical mechanics of a gas of free particles,
by following essentially the same lines as in \cite{IsakovIJMP94}.
A statistical distribution is derived, which again coincides
with the one for exclusion statistics \cite{Wu94,IsakovMPL94}.
We summarize the expressions for the cluster and virial coefficients
of a multispecies system, in the Appendix.

\section{Mutual fractional exchange statistics}

When particles are distinguishable, their wave functions have no
definite behavior with respect to their interchange, but one can
impose certain conditions that they should satisfy under a double
interchange, or, equivalently, a winding of one particle around
the other one. In three dimensions, as usually, wave functions
are single-valued, whereas in two dimensions a non-trivial
phase factor may arise. In fact, if there are $M$ species of
particles, one has an $M\times M$ matrix $\alpha_{ab}$
of statistical parameters,
such that the wave function of the whole system picks up a phase
\begin{itemize}
\item $\exp[i\pi\alpha_{aa}]$ under an {\it interchange\/} of two
particles of species $a$
\item $\exp[2i\pi\alpha_{ab}]$ under {\it winding} a particle of
species $a$ around a particle of species $b$
\end{itemize}
(provided, in both cases, that the complete closed path encloses
no other particles). The second condition, of course, holds
for $a=b$ as well, but the first one in this case is stronger.
Note that there is periodicity in $\alpha_{aa}$'s with period~2,
but in $\alpha_{ab}$'s with period~1.
Note also that since ``winding particle $a$ around particle $b$'' and
``winding particle $b$ around particle $a$'' is one and the same
operation, in the system of their center of mass, the matrix
$\alpha_{ab}$ must be symmetric.

The above condition on the wave function can also be understood within the
approach where statistics appear as kinematic properties of the
configuration space \cite{Leinaas77}. The configuration space for a system
containing $M$ species of identical particles can be constructed
in analogy with the case of a single species
\cite{Brekke91}.
Irreducible representations of the fundamental group of the
configuration space determine properties of the wave function under
exchange of particles. In \cite{Brekke91} it was found that
the fundamental group for the configuration space for a system with
$M$ species of identical particles is $B_{N_1\dots N_M}$, a
generalization of the braid group $B_N$.
One-dimensional irreducible representations of $B_{N_1\dots N_M}$
(corresponding to spinless particles) are labeled by $M$ parameters
associated with exchange of particles of the same species (which may be
identified with the above $\alpha_{aa}$) and $\frac12 M(M-1)$ parameters
associated with exchange of particles of different species (which may be
identified with the above $\alpha_{ab}$) \cite{Brekke91}.

One important observation is now in order.
A physical picture usually associated with anyons is
the model of charge-flux composites \cite{Wilczek82}, where
the origin
of the anyonic interchange phase is the Aharonov-Bohm
interaction of the charges and fluxes ascribed to
the particles.
In the multispecies case, each species $a$ would be characterized by a
charge $e_a$ and a flux
$\phi_a$. Now, if one writes the Hamiltonian in the regular gauge,
including
the relevant vector potentials, it turns out
that a singular gauge transformation
to the free Hamiltonian (anyon gauge)
is possible only if the equality
\begin{equation}
e_a \phi_b = e_b \phi_a
\label{1}
\end{equation}
holds, for each $a$ and $b$. In this case, one will have
\begin{equation}
\alpha_{ab} = e_a \phi_b / 2\pi,
\label{2}
\end{equation}
and the matrix $\alpha_{ab}$ is indeed symmetric.
Equations (\ref{1}) and (\ref{2}) together imply that
\begin{equation}
\alpha_{ab} = \pm \sqrt{\alpha_{aa}\alpha_{bb}}.
\label{2bis}
\end{equation}
Thus, only the diagonal statistical parameters are independent,
save for the sign.
However, a situation may be imagined when there are several
Aharonov-Bohm gauge fields. Eq.~(\ref{1}) then becomes
\begin{equation}
e_a^\beta \phi_b^\beta = e_b^\beta \phi_a^\beta,
\label{3}
\end{equation}
$\beta$ labeling the gauge fields, while (\ref{2})
becomes $\alpha_{ab} = \sum_\beta e_a^\beta \phi_b^\beta / 2\pi$,
and (\ref{2bis}) no longer holds.
Therefore, one may consider all $\alpha_{ab}$'s as independent,
but in the Aharonov-Bohm model this would, generally speaking,
demand to have several gauge fields.
Note at the same time that
in a model where the fluxes appear as a result of coupling the
particles to Chern-Simons gauge fields \cite{Dasnieres92},
(\ref{3}) is satisfied automatically.

Consider now the simplest case of $M=2$, and denote for brevity
$\alpha_{11}=\alpha_1$, $\alpha_{22}=\alpha_2$, $\alpha_{12}=\gamma$.
A many-particle wave function of a system of $N_1$ particles
of species~1 and $N_2$ particles of species~2 that satisfies the
above-discussed interchange conditions reads in the anyon gauge
\begin{equation}
\Psi =
\prod\limits_{j<k}^{N_1}(z_j-z_k)^{\alpha_1}
\prod\limits_{m<n}^{N_2}(\zeta_m-\zeta_n)^{\alpha_2}
\prod\limits_{j,m}(z_j-\zeta_m)^\gamma \Phi(\{z_j\},\{\zeta_m\})
\label{4}
\end{equation}
$\alpha_1,\alpha_2,\gamma>0$ is meant, while the replacement
$(z_j-z_k)^{\alpha_1} \to (z^*_j-z^*_k)^{-\alpha_1}$ is to be made
if $\alpha_1 < 0$, etc. $\Phi$ is a single-valued function symmetric with
respect to the
coordinates $\{z_j\}$ (spe\-cies~1) as well as with respect to the
coordinates $\{\zeta_m\}$ (species~2), but arbitrary with
respect to $z \leftrightarrow \zeta$. This is the starting point
for constructing the $(N_1+N_2)$-particle spectrum. Obviously, as
far as $\gamma$ is not an integer,
all $(N_1+N_2)$ coordinates are ``entangled'' together, while if
$\gamma$ is integer, the problem can be factorized into the one of
$N_1$ particles and the one of $N_2$ particles.

Knowing the spectrum, one may find the partition function
$Z_{N_1N_2}\! = \!{\rm Tr}\:\exp \left( -\beta H_{N_1N_2} \right)$,
then proceed to the grand partition function
\begin{equation}
\Xi = \sum_{N_1,N_2} z_1^{N_1} z_2^{N_2} Z_{N_1N_2},
\label{5}
\end{equation}
where $z_a = \exp(\beta\mu_a)$ are the fugacities, in order to obtain the
cluster expansion
\begin{equation}
\ln \Xi = \sum_{k_1,k_2} b_{k_1k_2} z_1^{k_1} z_2^{k_2}
\label{5bis}
\end{equation}
and the virial expansion for the
equation of state
\begin{equation}
\beta P = \sum_{k_1,k_2} a_{k_1k_2} \rho_1^{k_1} \rho_2^{k_2}
\label{6}
\end{equation}
(see Appendix).

Clearly, for the virial coefficients $a_{n0}$ and $a_{0n}$ one has
\begin{equation}
a_{n0} = a_n(\alpha_1), \qquad a_{0n} = a_n(\alpha_2),
\label{7}
\end{equation}
where $a_n(\alpha)$ is the $n$-th virial coefficient of anyons
with statistical parameter $\alpha$. Among these, only $a_2$ is
known exactly \cite{Arovas85},
\begin{equation}
a_2(\alpha) = \left[ \frac{1}{4} - \frac{1}{2}(1-\alpha)^2 \right]
\lambda^2,
\label{7a}
\end{equation}
with $\lambda=\sqrt{2\pi\beta/m}$.
The mixed second-order coefficient, $a_{11}$, involves
the $N_1=N_2=1$ case. Eq.~(\ref{4}) then turns into
\begin{equation}
\Psi = (z - \zeta)^\gamma \Phi(z,\zeta),
\label{8}
\end{equation}
where $\Phi$ is {\it any\/} single-valued function. Since it
may always be represented as a sum of symmetric and antisymmetric functions,
the partition function $Z_{11}$ can be written as
\begin{equation}
Z_{11}=Z_2(\gamma)+Z_{2}(1+\gamma),
\label{9a}
\end{equation}
$Z_{2}(\gamma)$ being the 2-anyon partition function
with statistical parameter $\gamma$. Using the formulas from the
Appendix, we get
\begin{equation}
a_{11}=a_{2}(\gamma)+a_{2}(1+\gamma) = \gamma(1-\gamma) \lambda^2.
\label{10}
\end{equation}
For $\gamma = 0,1$, this coefficient vanishes, reflecting the fact
that particles of different species are mutually independent.

\section{Quantum and statistical mechanics in the lowest Landau level}

Calculating higher-order virial coefficients would
demand solving a multiparticle problem, which is in general
impossible. In the one-species case, however,
-when the system is projected in the lowest
Landau level (LLL) of a strong external magnetic field,
the whole virial expansion
can be obtained\cite{Dasnieres94a}.
Note that for the gap above the ground state to be
bounded from below and thus the projection onto the LLL to
be justified, the particle fluxes must be antiparallel to
the external field, for example if $B<0$, one should have $\phi>0$.
If the particle charge $e$ is taken to be positive, then
the necessary condition is $\alpha \in [0,1]$ \footnote{Note that
this convention for signs is opposite to \cite{Dasnieres94a,Dasnieres94b}
and coincides with \cite{Wu94}.}.
The equation of state then is
\begin{equation}
\beta P = \rho_L \ln \left( 1 + \frac{\nu}{1 - \alpha\nu} \right)
\label{11}
\end{equation}
($\rho_L = m\omega_c / \pi$ is the Landau degeneracy per unit area,
$\rho = N/V$ is the density,
$\nu = \rho/\rho_L$ is the filling factor,
$\omega_c = |eB|/2m$ is half the cyclotron frequency;
``strong magnetic field'' means $\beta\omega_c \gg 1$).

It has been shown \cite{Wu94}
that the same equation (\ref{11}) is obtained from Haldane's
definition of fractional statistics \cite{Haldane91},
usually referred to as exclusion statistics, with
$\alpha$ being the quantity by which the number of available
single-particle states diminishes as one particle is added to
the system. Why the equations coincide can be understood within a mean-field
approximation (where it becomes possible to speak about single-particle
states): Since the anyon
fluxes are antiparallel to the external field,
the flux $\phi = 2\pi\alpha/e$ of a particle
added partially screens the external flux $\Phi = BV$
and thus diminishes the number of states available for other
particles; when $N$ particles are added, the Landau level degeneracy
per unit area becomes $\rho_L - \rho\alpha$.

A multispecies generalization of (\ref{11}) within the
framework of exclusion statistics is as follows:
If $\alpha_{ab}$ is the decrease in the number of single-particle
states available for $a$-th species caused by addition of one
particle of $b$-th species, then
\begin{equation}
\beta P = \rho_{L} \sum_a \ln
\left( 1 + \frac{\nu_a}{1 - \sum\limits_b \alpha_{ab}\nu_b} \right),
\label{12}
\end{equation}
with $\nu_a = \rho_a/\rho_{L}$ \cite{Dasnieres94b,Wu94}.

The physical picture would be the following: a particle
of $a$-th species carries statistical charges $e_a^\beta$,
which couple to the statistical gauge fields, and
an electric charge $e$ which
couples to the external magnetic field and
has to be common to all species. In fact, we are forced
to restrict ourselves to the Landau density $\rho_L$ being
the same for all species (whereas in \cite{Dasnieres94b,Wu94},
where mean field arguments are used,
different Landau densities $\rho_{La}$  are possible),
because it is only in this case that exact eigenstates
can be constructed.

In the two-species case,
by using microscopic arguments for the LLL spectrum,
we will now show
that the above equation indeed holds
with $\alpha_{ab}$'s being
the mutual statistical parameters defined previously:
it is then almost certain that (\ref{12}) is correct
for any number of species.

If there is only one statistical charge, then,
as already emphasized, $\gamma$ is not independent from the $\alpha_{aa}$'s.
Now, both $\phi_1$ and $\phi_2$ have to be positive, and
it follows then from (\ref{1}) that $e_1$ and $e_2$  must
have the same sign; choosing it again to be positive,
we have $\alpha_1, \alpha_2 \in [0,1]$, and, by
virtue of (\ref{2}), $\gamma$ is positive as well, so
the constraint (\ref{2bis}) would read
\begin{equation}
\gamma = \sqrt{\alpha_1\alpha_2}.
\label{SFO}
\end{equation}
However, our considerations will hold in the case of several
gauge fields as well,
and the result will be true for arbitrary $\gamma \in [0,1]$
(with the condition, however, that the gap above the ground
state be bounded from below by a quantity of the order the Landau gap
$2\omega_c$, which is certainly true for $\gamma$ small enough).

By analogy with \cite{Dasnieres94a}, we add a harmonic
attraction $\sum_{j=1}^{N_1+N_2} \frac{1}{2} m \omega^2 r_j^2$ and
observe, from (\ref{4}), that the spectrum of the
$(N_1+N_2)$-particle system in the LLL becomes
that of a system of $N_1$ and $N_2$ bosons
(mutually {\it independent\/}) in a
``harmonic-Landau'' potential whose single-particle
spectrum reads
\begin{equation}
\varepsilon_{\ell} = \omega_t + (\omega_t - \omega_c)\ell
\label{13}
\end{equation}
($\omega_t = \sqrt{\omega_c^2 + \omega^2}, \; \ell = 0,1,\ldots$),
plus a constant shift
\begin{equation}
\Delta E_{N_1N_2} (\alpha_1, \alpha_2, \gamma) =
\left[ \frac{N_1(N_1-1)}2\alpha_1 + \frac{N_2(N_2-1)}2\alpha_2
+ N_1N_2\gamma \right] (\omega_t - \omega_c)
\label{14}
\end{equation}
coming from the anyonic prefactor in (\ref{4}).
One then has
\begin{equation}
Z_{N_1N_2} = e^{-\beta \Delta E_{N_1N_2} (\alpha_1,\alpha_2,\gamma)}
Z^b_{N_1} Z^b_{N_2},
\label{15}
\end{equation}
where
\begin{equation}
Z^b_N = \frac{e^{-N\beta\omega_t}}
{(1-e^{-\beta(\omega_t - \omega_c)})(1-e^{-2\beta(\omega_t - \omega_c)})
\cdots(1-e^{-N\beta(\omega_t - \omega_c)})}
\label{16}
\end{equation}
is the $N$-boson partition function in the harmonic-Landau potential.

For the cluster coefficients, one gets (see Appendix for the
prescription for passing to the thermodynamic limit)
\begin{eqnarray}
b_{k_10} & \!\!\!=\!\!\! & \rho_LV \frac{1}{k_1}
\prod_{l_1=1}^{k_1-1} \left( 1 - \frac{k_1\alpha_1}{l_1} \right) \cdot
e^{-k_1\beta\omega_c}, \label{17} \\
b_{k_1k_2} & \!\!\!=\!\!\! & -\rho_LV\gamma \frac{(k_1+k_2)}{k_1k_2}
\prod_{l_1=1}^{k_1-1}\left( 1 - \frac{k_1\alpha_1 + k_2\gamma}{l_1} \right)
\prod_{l_2=1}^{k_2-1}\left( 1 - \frac{k_2\alpha_2 + k_1\gamma}{l_2} \right)
\cdot
e^{-(k_1+k_2)\beta\omega_c},
\nonumber \\ \label{18}
\end{eqnarray}
and $b_{0k_2} = b_{k_20} |_{\alpha_1 \leftrightarrow \alpha_2}$.
Note that, as  expected,  (\ref{18}) does not reduce to (\ref{17}) when $k_2=0$
(in particular, $b_{k_1k_2}$ certainly vanishes for $\gamma=0$,
while $b_{k_10}$ is $\gamma$ independent). The virial coefficients
calculated herefrom take the form ($\beta\omega_c \gg 1$)
\begin{equation}
a_{k_1k_2} = -\frac{1}{\rho_L^{k_1+k_2-1}}
\frac{(k_1+k_2-1)!}{k_1!k_2!}
\left\{ [(\alpha_1-1)^{k_1} - \alpha_1^{k_1}] \gamma^{k_2}
 +      [(\alpha_2-1)^{k_2} - \alpha_2^{k_2}] \gamma^{k_1}
\right\},
\label{19}
\end{equation}
and the equation of state then is
\begin{equation}
\beta P = \rho_L \left[
\ln \left( 1 + \frac{\nu_1}{1 - \alpha_1\nu_1 - \gamma\nu_2} \right) +
\ln \left( 1 + \frac{\nu_2}{1 - \alpha_2\nu_2 - \gamma\nu_1} \right)
\right],
\label{20}
\end{equation}
which indeed coincides with (\ref{12}).

A mean-field interpretation is possible in this case as well.
Indeed, what figures in the Hamiltonian
for $a$-th species is the sum $e\vec A + \sum_\beta e^\beta_a \vec A^\beta$,
where $\vec A$ is the vector potential of the magnetic field and
$\vec A^\beta$ is the $\beta$-th gauge field potential created
by all other particles.
If {\it all\/} the gauge fields are averaged,
so that $B^\beta = \sum_b N_b \phi^\beta_b /V = \sum_b \rho_b \phi^\beta_b$,
then the Landau level degeneracy per unit area becomes
$\frac{1}{2\pi} (|eB| - \sum_\beta e_a^\beta B^\beta) = \rho_L
- \sum_b \rho_b \alpha_{ab}$, which is precisely what leads to (\ref{12}).

As well as for the one-species case, this equation is valid only
when $\nu_1$ and $\nu_2$ are small enough---in fact,
both denominators have to be positive;
one of them vanishing means that the critical filling is
achieved for the corresponding species. In the special case
(\ref{SFO}), the condition is
\begin{equation}
\sqrt{\alpha_1}\nu_1 + \sqrt{\alpha_2}\nu_2
< \frac{1}{\sqrt{\max (\alpha_1,\alpha_2) }}.
\end{equation}

\section{Interpretation in terms of single-state distributions}

We now wish to interpret the exactly
solvable model of anyons in the LLL in
terms of a gas of free particles with a particular
statistical distribution. We will use arguments similar
to those of \cite{IsakovIJMP94} for the Calogero model.
Description of single-state partition functions for multispecies systems
follows Ref. \cite{IsakovMPL94}.

Let there be single-particle states labeled with $i$, common for
all species $a$, but with energies
$\varepsilon_a^{(i)}$ possibly depending on $a$. The particles
in different single-particle states are assumed to be
statistically independent, which implies that the grand partition
function is a product of single-state grand partition functions
\begin{equation}
\Xi =\prod_i \Xi^{(i)}
\label{eq:a1}\end{equation}
with $\Xi^{(i)}$ depending only on the Gibbs factors
$x_a=e^{\beta(\mu_a-\varepsilon_a^{(i)})}$:
$\Xi^{(i)}=\Xi^{(i)}(\{x_a\})$.
The statistical distributions (average number of
particles of different species in state $i$) are
\begin{equation}
n_a^{(i)}=x_a\frac{\partial}{\partial x_a}\ln \Xi^{(i)}.
\label{eq:a2}\end{equation}
We assume that $\ln \Xi^{(i)}$ may be expanded in integer powers
of the Gibbs factors (again, we specify the
formulas containing the series expansions for two species):
\begin{equation}
\ln \Xi^{(i)} =\sum_{k_1,k_2=0}^\infty f_{k_1k_2}x_1^{k_1}x_2^{k_2}.
\label{eq:a3}\end{equation}
It follows then from (\ref{eq:a2})
\begin{equation}
n_a^{(i)}=\sum_{k_1,k_2=1}^\infty k_a f_{k_1k_2}x_1^{k_1}x_2^{k_2}.
\label{eq:a4}\end{equation}

Combining (\ref{eq:a1}) with (\ref{eq:a3}), summing over
single-particle states and comparing with (\ref{5bis}),
we get the relation
\begin{equation}
b_{k_1k_2}=f_{k_1k_2} \sigma_{k_1k_2},
\label{eq:a6}\end{equation}
where
\begin{equation}
\sigma_{k_1k_2}=\sum_i
e^{-\beta(k_1\varepsilon_1^{(i)}+k_2\varepsilon_2^{(i)})}.
\label{eq:a7}\end{equation}

We use the many-particle partition functions for anyons in the LLL in a
harmonic well (\ref{15}), with $\omega\to0$,
to calculate the coefficients $f_{k_1k_2}$ in the
expansion (\ref{eq:a3}). Expressing the cluster coefficients
in terms of the many-particle partition functions
(but {\it not yet\/} using the thermodynamic limit prescription given
in the Appendix),
taking into account the expression (\ref{13}) for the single-particle
spectrum, in order to calculate $\sigma_{k_1k_2}$,
and then using (\ref{eq:a7}) and (\ref{eq:a6}), we eventually obtain
\begin{eqnarray}
f_{k_10} &\!\!\!=\!\!\!& \frac1{k_1}
\prod_{l_1=1}^{k_1-1}(1-\frac{k_1\alpha_1}{l_1}),
\label{eq:a8} \\
f_{k_1k_2} &\!\!\!=\!\!\!& -\gamma\frac{k_1+k_2}{k_1k_2}
\prod_{l_1=1}^{k_1-1}(1-\frac{k_1\alpha_1+k_2\gamma}{l_1})
\prod_{l_2=1}^{k_2-1}(1-\frac{k_2\alpha_2+k_1\gamma}{l_2}).
\label{eq:a9}
\end{eqnarray}

We thus conclude that in the limit $\omega\to 0$, multispecies
anyons in the LLL can be described by
the partition functions (\ref{eq:a3}) and the
statistical distributions (\ref{eq:a4}) with the coefficients of their
expansions in powers of the Gibbs factors given by
(\ref{eq:a8})--(\ref{eq:a9}).

Now we consider directly
the case $\omega=0$. All the single-particle states have then
the same energy $\omega_c$, therefore
all the relevant quantities do not depend on the index $i$ any longer,
and summation over single-particle states
reduces to multiplication by the number of states in the LLL,
$\rho_L V$, e.~g.~$\ln\Xi = \rho_L V \ln\Xi^{(1)}$.
It turns out then that $\Xi^{(1)}$ can be represented as
\begin{equation}
\Xi^{(1)}=\prod_a\Xi_a
\label{eq:a10}\end{equation}
with
\begin{equation}
\Xi_a=1+\frac{n_a}{1-\sum_b \alpha_{ab}n_b},
\label{eq:a11}\end{equation}
so that for the coefficients of the expansions of the
single-state grand partition functions
\begin{equation}
\Xi_a =\sum_{k_1,k_2} P_{a;k_1k_2}x_1^{k_1}x_2^{k_2}
\label{eq:a12}\end{equation}
we obtain
\begin{eqnarray}
P_{1;k_10} &\!\!\!=\!\!\!& \prod_{l_1=2}^{k_1}
\left( 1 - \frac{k_1\alpha_1}{l_1} \right), \label{eq:Po} \\
P_{1;k_1k_2} &\!\!\!=\!\!\!& -\gamma\frac{k_1}{k_2}
\prod_{l_1=2}^{k_1}\left(1-\frac{k_1\alpha_1+k_2\gamma}{l_1}\right)
\prod_{l_2=1}^{k_2-1}\left(1-\frac{k_2\alpha_2+k_1\gamma}{l_2}\right)
\label{eq:Pg}\end{eqnarray}
and $P_{2;k_1k_2}=P_{1;k_2k_1}|_{\alpha_1\leftrightarrow\alpha_2}$.

Eqs.~(\ref{eq:a8}) and (\ref{eq:Po}) have been previously derived
for fractional statistics in one
dimension, starting from the Calogero model
\cite{IsakovIJMP94} (see also a
recent derivation directly for exclusion statistics \cite{Poly}).
Eqs.~(\ref{eq:a9}) and (\ref{eq:Pg}) are
generalizations onto the multispecies case.

The partition functions $\Xi_a$, in addition, satisfy the equations
\begin{equation}
\Xi_a=1+x_a\prod_b\Xi_b^{\delta_{ab}-\alpha_{ab}}
\label{eq:a13}\end{equation}
Substituting (\ref{eq:a11}) into (\ref{eq:a13}), we get the equation for
the statistical distributions $n_a$
\begin{equation}
n_a/x_a=\prod_b (1-\sum_c \alpha_{bc}n_c)^{\alpha_{ab}}
(1-n_a-\sum_c \alpha_{bc}n_c)^{\delta_{ab}-\alpha_{ab}}.
\label{eq:a14}\end{equation}

These statistical distributions coincide with the
ones for exclusion statistics \cite{Wu94,IsakovMPL94}
after the identification
\begin{equation}
g_{ab}=\alpha_{ab}
\label{eq:a18}\end{equation}
(in our model, consequently, the exclusion
statistics parameters are  symmetric, $g_{ab}=g_{ba}$).

It is now easy to calculate the cluster coefficients,
using (\ref{eq:a6}). Using the thermodynamic
limit prescription at each order $k_1+k_2$, (\ref{eq:a7}) becomes
\begin{equation}
\sigma_{k_1k_2}=\rho_L V e^{-\beta(k_1+k_2)\omega_c}
\label{eq:a19}
\end{equation}
and, with the use of (\ref{eq:a8})--(\ref{eq:a9}),
the formulas (\ref{17})--(\ref{18}) are
recovered.
Note finally that once (\ref{eq:a10}) and (\ref{eq:a11}) have been
correctly ``guessed'', the equation of
state is nothing but the thermodynamic identity $\beta P V = \ln \Xi$.

\section{Concluding remarks}

We have studied quantum and statistical mechanics for mutual fractional
exchange statistics. We have presented a microscopic solvable model, anyons
in the LLL, which, on one hand, involves mutual fractional exchange
statistics and, on the other hand, reproduces the statistical mechanics
for mutual fractional exclusion statistics. This establishes correspondence
between the two approaches to mutual statistical interactions.

We hope that due to the similarity between anyons in the LLL and FQHE
quasi\-part\-ic\-les, the present work will help a better understanding of
excitations in quantum Hall systems.

In addition, intimate links between anyons in the LLL and the Calogero
model suggest that the microscopic solutions for multispecies
anyons found in this paper can be used to construct solutions for a
multispecies Calogero model which would reveal mutual exclusion statistics
\cite{furt}.

\bigskip
{\bf Acknowledgements:}
We thank K\aa re Olaussen for valuable discussions. We are also grateful
to NORDITA for their hospitality during the Anyon Workshop where this
work was initiated.
S.M. gratefully acknowledges warm
hospitality of the theory division of the IPN at Orsay
where a significant part of this work was done and thanks
Guillermo Zemba for bringing ref.~\cite{Froelich91} to his attention.
S.B.I. was supported in part by the Russian Foundation for Fundamental
Research under grant No.~95-02-04337.

\bigskip
{\Large \bf Appendix}
\bigskip
\nopagebreak
\setcounter{equation}{0}
\renewcommand{\theequation}{A.\arabic{equation}}
\newcommand{\bt}{\tilde{b}}
\newcommand{\at}{\tilde{a}}

We summarize here, for reference purposes, the formulas
concerning the cluster and virial expansions for a two-species
gas, and give the relations between $Z_{N_1N_2}$, $b_{n_1n_2}$,
and $a_{n_1n_2}$, up to the fifth order.

The grand canonical partition function is
\begin{equation}
\Xi = \sum_{N_1,N_2} z_1^{N_1} z_2^{N_2} Z_{N_1N_2}
\label{A1}
\end{equation}
and the cluster coefficients $b_{k_1k_2}$ are determined
in terms of $Z_{N_1N_2}$'s by writing the logarithm of (\ref{A1})
in the form
\begin{equation}
\ln \Xi = \sum_{k_1,k_2} b_{k_1k_2} z_1^{k_1} z_2^{k_2}.
\label{A2}
\end{equation}
The pressure and the densities are
\begin{equation}
\beta P = \frac{1}{V} \ln \Xi, \qquad \rho_a = \frac{z_a}{V}
\frac{\partial \ln \Xi}{\partial z_a},
\label{A3}
\end{equation}
respectively ($V$ is the area); writing
\begin{equation}
\beta P = \sum_{k_1,k_2} a_{k_1k_2} \rho_1^{k_1} \rho_2^{k_2},
\label{A4}
\end{equation}
using (\ref{A3}) with $\ln \Xi$ substituted from (\ref{A2}),
and matching coefficients at equal powers of $z$'s on both sides,
one gets the equations which allow to express
$a_{k_1k_2}$ in terms of $b_{k_1k_2}$'s.

The explicit formulas are given below. For the cluster coefficients, one has
\begin{eqnarray*}
b_{10} & = & Z_{10}; \\
\\
b_{20} & = & -\frac{1}{2}Z_{10}^2 + Z_{20}, \\
b_{11} & = & -Z_{01}Z_{10} + Z_{11}; \\
\\
b_{30} & = & \frac{1}{3}Z_{10}^3 - Z_{10}Z_{20} + Z_{30}, \\
b_{21} & = & Z_{01}Z_{10}^2 - Z_{10}Z_{11} - Z_{01}Z_{20} + Z_{21}; \\
\\
b_{40} & = & -\frac{1}{4}Z_{10}^4 + Z_{10}^2Z_{20} - \frac{1}{2}Z_{20}^2
             - Z_{10}Z_{30} + Z_{40}, \\
b_{31} & = & -Z_{01}Z_{10}^3 + Z_{10}^2Z_{11} + 2Z_{01}Z_{10}Z_{20}
             -Z_{11}Z_{20} - Z_{10}Z_{21} - Z_{01}Z_{30} + Z_{31}, \\
b_{22} & = & -\frac{3}{2}Z_{01}^2Z_{10}^2 + Z_{02}Z_{10}^2 +
             2Z_{01}Z_{10}Z_{11} - \frac{1}{2}Z_{11}^2 - Z_{10}Z_{12}
             + Z_{01}^2Z_{20} \\
          && {} - Z_{02}Z_{20} - Z_{01}Z_{21} + Z_{22}; \\
\\
b_{50} & = & \frac{1}{5}Z_{10}^5 - Z_{10}^3Z_{20} + Z_{10}Z_{20}^2
             + Z_{10}^2Z_{30} - Z_{20}Z_{30} - Z_{10}Z_{40} + Z_{50}, \\
b_{41} & = & Z_{01}Z_{10}^4 - Z_{10}^3Z_{11} - 3Z_{01}Z_{10}^2Z_{20}
             + 2Z_{10}Z_{11}Z_{20} + Z_{01}Z_{20}^2 + Z_{10}^2Z_{21} \\
          && {} - Z_{20}Z_{21} + 2Z_{01}Z_{10}Z_{30} - Z_{11}Z_{30}
             - Z_{10}Z_{31} - Z_{01}Z_{40} + Z_{41}, \\
b_{32} & = & 2Z_{01}^2Z_{10}^3 - Z_{02}Z_{10}^3 - 3Z_{01}Z_{10}^2Z_{11}
             + Z_{10}Z_{11}^2 + Z_{10}^2Z_{12} - 3Z_{01}^2Z_{10}Z_{20} \\
          && {} + 2Z_{02}Z_{10}Z_{20} + 2Z_{01}Z_{11}Z_{20} - Z_{12}Z_{20}
             + 2Z_{01}Z_{10}Z_{21} - Z_{11}Z_{21} \\
          && {} - Z_{10}Z_{22} + Z_{01}^2Z_{30} - Z_{02}Z_{30}
             - Z_{01}Z_{31} + Z_{32}.
\end{eqnarray*}
If a harmonic oscillator regularization is used,
the thermodynamic limit is meant as $\omega\to0$, with a
particular prescription at each order of the cluster expansion \cite{McCabe}.
A treatment analogous to the one given in
\cite{Olaussen92} shows that to get the correct thermodynamic limit
for $b_{k_1k_2}$, the necessary prescription is
to identify $\frac{2\pi}{(k_1+k_2)\beta m \omega^2} \to V$, in the
limit $\omega \to 0$. For the virial coefficients, defining
\begin{equation}
\tilde{b}_{k_1k_2} = b_{k_1k_2} / b_{10}^{k_1} b_{01}^{k_2},
\end{equation}
\begin{equation}
\tilde{a}_{k_1k_2} = a_{k_1k_2} / V^{k_1+k_2-1},
\end{equation}
one has
\begin{eqnarray*}
\at_{10} & = & 1;\\
\\
\at_{20} & = & -\bt_{20},\\
\at_{11} & = & -\bt_{11};\\
\\
\at_{30} & = & 4\bt_{20}^2 - 2\bt_{30},\\
\at_{21} & = & \bt_{11}^2 + 4\bt_{11}\bt_{20} - 2\bt_{21};\\
\\
\at_{40} & = & -20\bt_{20}^3 + 18\bt_{20}\bt_{30} - 3\bt_{40}, \\
\at_{31} & = & -\bt_{11}^3 - 6\bt_{11}^2\bt_{20} - 24\bt_{11}\bt_{20}^2
               + 3\bt_{11}\bt_{21} + 12\bt_{20}\bt_{21} + 9\bt_{11}\bt_{30}
               - 3\bt_{31}, \\
\at_{22} & = & -9\bt_{02}\bt_{11}^2 - 3\bt_{11}^3 + 6\bt_{11}\bt_{12}
               - 12\bt_{02}\bt_{11}\bt_{20} - 9\bt_{11}^2\bt_{20}
               + 6\bt_{12}\bt_{20} + 6\bt_{02}\bt_{21} \\
             && {} + 6\bt_{11}\bt_{21} - 3\bt_{22}; \\
\\
\at_{50} & = & 112\bt_{20}^4 + 32\bt_{40}\bt_{20} + 18\bt_{30}^2
               - 144\bt_{30}\bt_{20}^2 - 4\bt_{50}, \\
\at_{41} & = & \bt_{11}^4 + 8\bt_{11}^3\bt_{20} + 40\bt_{11}^2\bt_{20}^2
               + 160\bt_{11}\bt_{20}^3 - 4\bt_{11}^2\bt_{21} -
               24\bt_{11}\bt_{20}\bt_{21} - 80\bt_{20}^2\bt_{21}\\
            && {} + 2\bt_{21}^2 - 12\bt_{11}^2\bt_{30} - 120\bt_{11}\bt_{20}
               \bt_{30} + 24\bt_{21}\bt_{30} + 4\bt_{11}\bt_{31}
               + 24\bt_{20}\bt_{31} \\
            && {} + 16\bt_{11}\bt_{40} - 4\bt_{41}, \\
\at_{32} & = & 16\bt_{02}\bt_{11}^3 + 6\bt_{11}^4 - 12\bt_{11}^2\bt_{12}
               + 48\bt_{02}\bt_{11}^2\bt_{20} + 32\bt_{11}^3\bt_{20}
               - 32\bt_{11}\bt_{12}\bt_{20} \\
            && {} + 64\bt_{02}\bt_{11}\bt_{20}^2 + 80\bt_{11}^2\bt_{20}^2
               - 32\bt_{12}\bt_{20}^2 - 24\bt_{02}\bt_{11}\bt_{21}
               - 20\bt_{11}^2\bt_{21} + 8\bt_{12}\bt_{21} \\
            && {} - 32\bt_{02}\bt_{20}\bt_{21} - 64\bt_{11}\bt_{20}\bt_{21}
               + 8\bt_{21}^2 + 8\bt_{11}\bt_{22} + 16\bt_{20}\bt_{22}
               - 24\bt_{02}\bt_{11}\bt_{30} \\
            && {} - 24\bt_{11}^2\bt_{30} + 12\bt_{12}\bt_{30}
               + 8\bt_{02}\bt_{31} + 12\bt_{11}\bt_{31} - 4\bt_{32};
\end{eqnarray*}
with the obvious symmetry with respect to permutation of the
subscripts.


\begin{thebibliography}{99}
\bibitem{Leinaas77} J.M.~Leinaas and J.~Myrheim,
   Nuovo Cim. {\bf 37B} (1977) 1.
\bibitem{Wilczek82} F.~Wilczek, Phys.~Rev.~Lett. {\bf 48} (1982) 1144;
   {\it ibid.} {\bf 49} (1982) 957.
\bibitem{Mor} P.W.~Anderson, Science {\bf 235} (1987) 1196;
   G.~Baskaran et al., Solid State Commun.~{\bf 63} (1987) 973;
   T.~Mor, Phys.~Lett.~A {\bf 184} (1993) 29.
\bibitem{Wilczek92} F.~Wilczek, Phys. Rev. Lett. {\bf 69} (1992) 132;
   Z.~F.~Ezawa and A. Iwazaki, Phys. Rev. B {\bf 47} (1993) 7295 and
   references therein.
\bibitem{Froelich91} J.~Fr\"ohlich, A.~Zee, Nucl.~Phys. {\bf B364}
   (1991) 517.
\bibitem{Dasnieres94a} A.~Dasni\`eres de Veigy and S.~Ouvry,
   Phys.~Rev.~Lett. {\bf 72} (1994) 600.
\bibitem{Dasnieres94b} A.~Dasni\`eres de Veigy and S.~Ouvry,
   Mod.~Phys.~Lett.~B {\bf 9} (1995) 271.
\bibitem{Wu94} Y.-S.~Wu, Phys.~Rev.~Lett. {\bf 73} (1994) 922.
\bibitem{Haldane91} F.D.M.~Haldane, Phys.~Rev.~Lett. {\bf 67} (1991) 937.
\bibitem{IsakovIJMP94} S.B.~Isakov,
   Int.~J.~Mod.~Phys.~A {\bf 9} (1994) 2563.
\bibitem{IsakovMPL94} S.B.~Isakov, Mod.~Phys.~Lett.~B {\bf 8} (1994) 319.
\bibitem{Brekke91} L.~Brekke, A.F.~Falk, S.J.~Hugens and T.D.~Imbo,
   Phys.~Lett.~A {\bf 271} (1991) 73.
\bibitem{Dasnieres92} A.~Dasni\`eres de Veigy and S.~Ouvry,
   Phys.~Lett. {\bf B307} (1992) 91.
\bibitem{Arovas85} D.P.~Arovas, J.R.~Schrieffer, F.~Wilczek, A.~Zee,
   Nucl.~Phys. {\bf B251} (1985) 117.
\bibitem{Poly} A.~Polychronakos, preprint UUITP-03/95, hep-th/9503077.
\bibitem{furt} C. Furtlehner and S. Ouvry, IPNO/TH 94-02, to be
published in MPLB
\bibitem{McCabe} A.~Comtet, Y.~Georgelin and S.~Ouvry,
   J.~Phys.~A (Math.Gen.) {\bf 22} (1989) 3917;
   J.~McCabe and S.~Ouvry, Phys.~Lett. {\bf B260} (1991) 113.
\bibitem{Olaussen92} K.~Olaussen, Trondheim preprint No.~13 (1992).

\end{thebibliography}
\end{document}